\documentstyle[12pt]{article}
\begin{document}
\thispagestyle{empty}
\begin{center}
{\bf\Large Supergravity before 1976}\footnote{Work supported in part by
the ISF Grant No RY 9000 and INTAS Grant No 93--493.}

\bigskip
{\Large Dmitrij V. Volkov}

\bigskip
Kharkov Institute of Physics and Technology, Kharkov, 310108, Ukraine
\end{center}

\vspace{0.5cm}
\centerline{Abstract}
A story on how an attempt to realize  W. Heizenberg idea that the
neutrino might be a Goldstone particle had led in its development to the
discovery of supergravity action.
 \vspace{0.5cm}
 \begin{enumerate}
\item Introduction
\item Goldstone fermions, super--Poincar\'e group and
its spontaneous breaking
\item Gauged super--Poincar\'e group
\item
Supergravity action and super--Higgs effect
\item Superspace formulation
of Supergravity (first steps into Superspace)
\item Resum\'e
\end{enumerate}
\section{Introduction}

There are two kinds of fields which are directly related to continuous
symmetries.
These are the Goldstone and Yang--Mills fields.

The Goldstone fields are a manifestation of spontaneously broken
symmetries. The  Yang -- Mills fields appear when a symmetry is localized.

The both kinds of fields were predicted theoretically, and at the moment
of their invention seemed to have no relation to reality.

However, revealed by Higgs and others, their mutual interplay
exhibiting itself as the Higgs effect has thoroughly changed the situation.
And nowadays the concept of the gauge fields and of the
spontaneously broken symmetries is the backbone of theories unifying all
known interactions.

\bigskip
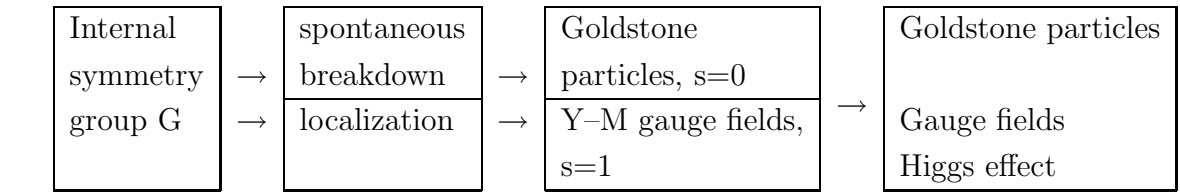
\begin{tabular}{|l|l|l|l|l|l|l|}
\cline{1-1}\cline{3-3}\cline{5-5}\cline{7-7}
 Internal & & spontaneous & &
 Goldstone &
& Goldstone~particles\\
symmetry &$\rightarrow$ & breakdown &$\rightarrow$ & particles, s=0& &\\
\cline{3-3}\cline{5-5}
group G &$\rightarrow$& localization&$\rightarrow$ &
 Y--M~gauge~fields,&\raisebox{1.2ex}{$\rightarrow$}& Gauge
fields\\
& & &&s=1&& Higgs effect\\ \cline{1-1}\cline{3-3}\cline{5-5}\cline{7-7}
\end{tabular}

\bigskip
\bigskip
The modern idea of the unification involves not only internal symmetries
but also a new type of symmetry discovered at the beginning of 70's,
namely, supersymmetry, which is intrinsically intertwines with the former.

Main features of the supersymmetry have been discussed in the talks by
P. Dimopolous and S. Fayet \cite{01}. So, I shall remind only that
supersymmetry was independently discovered by three groups of
authors:
\begin{center} Yu. Gol'fand \& E. Lichtman      (1971)\\
D. Volkov \& V. Akulov           (1972) \\
J. Wess \& B. Zumino             (1974) \end{center}

The motivation and starting points used by these groups
were quite different.

The motivation of Gol'fand and Lichtman \cite{1} was to introduce a parity
violation in the quantum field theory. The starting point of the paper by
Akulov and myself \cite{2,3} was the question whether Goldstone particles
with spin one half might exist.
Wess and Zumino \cite{4} performed the generalization of the supergroup
which first appeared in the Neveu--Schwarz--Ramond dual model \cite{5,6}.

The approach of the Volkov--Akulov papers is the most appropriate for
gauging the super--Poincar\'e group, which was done a little bit later in
the papers by Soroka and myself \cite{7,8}. The
last--mentioned papers are the
natural continuation of the Volkov--Akulov papers since the transformation
laws for gauge fields are determined by the same group structure as that
used for the description of the Goldstone fields.

As it has already been mentioned, our approach was caused by the question
about possible existence of the spin 1/2 Goldstone particles.

The second question closely related to the first one is: what is the group
whose breakdown yields Goldstone fields with the required properties?
 In  some respect this is the reverse of the usual consideration of
 Goldstone and gauge fields when an internal symmetry group is given and
the properties of the fields are derived from the known symmetry group
properties.

\bigskip
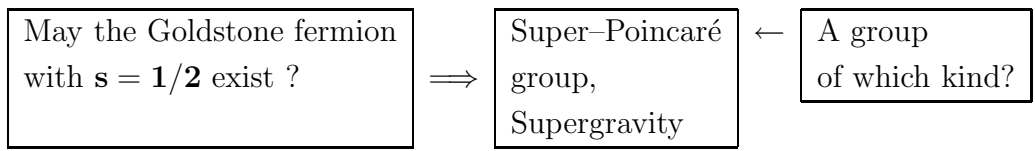
\begin{tabular}{|l|l|l|l|l|}\cline{1-1}\cline{3-3}\cline{5-5}
May the Goldstone fermion & &  Super--Poincar\'e &$\leftarrow$&
A group \\
with ${\bf s=1/2}$ exist ? &  $\Longrightarrow$ &group,&&
of which kind?  \\  \cline{5-5}
  & & Supergravity\\ \cline{1-1}
\cline{3-3} \end{tabular}

\bigskip
\bigskip
Answering these questions led us to the discovery of the super--Poincar\'e
group.

It is convenient to consider the internal symmetry groups and the
super--Poincar\'e group in parallel since it gives the possibility of
learning which features of the both cases are common and which are
different.

\bigskip
\bigskip
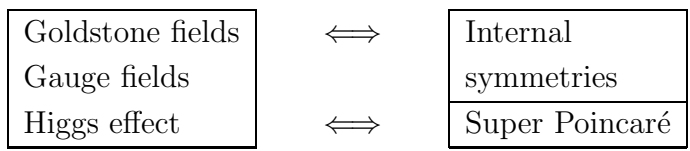
\begin{tabular}{|l|l|l|}\cline{1-1}\cline{3-3}
Goldstone fields &~~~~~$\Longleftrightarrow$~~~~~ & Internal \\
Gauge fields & & symmetries \\ \cline{3-3} \cline{3-3}
Higgs effect & ~~~~~$\Longleftrightarrow$~~~~~ &Super Poincar\'e \\
\cline{1-1} \cline{3-3}
\end{tabular}

\bigskip
\bigskip
As we shall see below the analogy between the two cases gives rather
rich intuitive insight which helps one to trivialize the procedure of
gauging the super--Poincar\'e group and getting the supergravity action.

Let us firstly recall that one of the main directions of the research in
theoretical particle physics in the 60's was the Current Algebra
initiated by M. Gell--Mann \cite{9,10,11}, and the PCAC hypothesis
\cite{101}, which produced many interesting results in the field of
weak and strong interactions, including a number of the so called soft
pion, or low energy, theorems (Y. Nambu \& D. Lurie \cite{12}, S. Adler
\cite{13}, S.  Weinberg \cite{14},...).  The essential progress was
achieved when this direction was goldstonized (Goldstone \cite{15}, Nambu,
Jona--Lozinio \cite{16}), and especially when S. Weinberg \cite{17} and
J. Schwinger \cite{18} proposed the method of Phenomenological Lagrangians
which easily reproduced all soft pion theorems with any number of pions.

At the time of the XIVth Conference on High Energy Physics (Vienna, 1968)
the problem of the Current Algebra and of the Phenomenological Lagrangians
was intensively discussed (see Weinberg's rapporteur talk \cite{20}).
There were two papers presented in the current algebra section of the
conference in which the generalization of the method of Phenomenological
Lagrangians to an arbitrary internal symmetry group had been elaborated.
One paper was presented by B. Zumino \cite{21} (co--authors C. Callan, S.
Coleman and J. Wess), and another one by myself \cite{22}. The main
results of the papers were practically identical. The difference was that
in Ref. \cite{22} \footnote{The papers \cite{22,24} are contained in
\cite{25}.} the works of E. Cartan on symmetric spaces and his method of
the exterior differential forms were intensively used.

The general procedure for constructing Phenomenological Lagrangians for
Goldstone particles of internal symmetry groups \cite{21,22} is
the following.

A group G is factorized
\begin{equation}\label{1}
G=KH,
\end{equation}
where H is a stationary subgroup of G, and K is the coset space K=G/H.

The $g$--algebra valued differential 1--forms
$$
G^{-1}dG=H^{-1}(K^{-1}dK)H + H^{-1}dH
$$
represent the vielbeins $(G^{-1}dG)_k$ and the connection 1--forms
$(G^{-1}dG)_h$, where $k$ and $h$ are subspaces of the g--algebra
corresponding, respectively, to $K$ and $H$.

The Phenomenological Lagrangian is constructed out of the vielbeins:
\begin{equation}\label{01}
L={1\over 2}Sp(G^{-1}dG)_k(G^{-1}dG)_k,
\end{equation}
while the connection forms are used to include interaction with other
particles which belong to representations of the group $G$.
G--multiplets are reduced to H--multiplets
with respect to the stationary subgroup H and considered as independent in
the approach in question. The splitting of the G--multiplets into the
H--multiplets together with the appearance of Goldstone fields is the main
feature of the Phenomenological Lagrangian method.

A little bit later this procedure was generalized by J. Wess and B. Zumino
\cite{23} by introducing the so called Wess--Zumino term, and by me for
the case of spontaneously broken symmetry groups containing the Poincar\'e
group as a subgroup \cite{24,25}\footnote{In \cite{24}, which was
prepared in autumn, 1971 there is the first mentioning of the generalization
of the Poincar\'e group to the super--Poincar\'e group.}.  The
Phenomenological Lagrangian for Goldstone fermions, which we now turn to
consider, is an example of the latter generalization.

\section{Goldstone Fermions, super--Poincar\'e group and its spontaneous
breaking}

Now we can return to the question how one can generalize the Poincar\'e
group to the super--Poincar\'e group by requiring Goldstone particles to
have spin one half. As it has been explained
in the previous section, the quantum numbers of the Goldstone particles
coincide with the quantum numbers of the K--generators.
Therefore, to ensure the appearance of the Goldstone fermions, the
Poincar\'e group should be generalized in such a way that the generalized
version contains generators with spin one--half obeying commutation
relations corresponding to the Fermi statistics. From a technical point of
view the problem was what representation of the Poincar\'e group is the
most appropriate for such  generalization.

The solution to this technical problem was that the following
representation of the Poincar\'e group had all required properties:
\begin{equation}\label{2}
{\bf G_{Poincare}=\left(
\begin{array}{cc}
1&iX\\
0&1
\end{array}\right)
\left(
\begin{array}{cc}
L&0\\
0&(L^+)^{-1}
\end{array}\right)},
\end{equation}
where $X_{\alpha\dot\beta}$, $L_{\alpha}^{~\beta}$ and
$\bf{(L^+)^{-1}}^{\dot\alpha}_{~\dot\beta}$ are $\bf{2\times 2}$ matrices.

In the generalization to the super--Poincar\'e group $K_{transl.}$ plays
the main role.

 Let us write it down in a form consisting of four blocks:
 \begin{equation}\label{3}
K=\left(
\begin{array}{cc}
1&iX\\
0&1
\end{array}\right)
\end{equation}

 Separating the blocks as follows
 \begin{equation}\label{4}
K^\prime=\left(\begin{array}{ccc}
1&-&i X\\
0&1&-\\
0&0&1
\end{array}\right)
 \end{equation}
 one can insert into the newly formed hatched blocks Grassmann spinors
 $\theta_\alpha$ and $\bar\theta_{\dot\alpha}$ so that $K^\prime$ becomes
 \begin{equation}\label{5}
 K^\prime=\left(\begin{array}{ccc}
1&\theta&i X^\prime\\
0&1&\bar\theta\\
0&0&1
\end{array}\right).
 \end{equation}

 The matrices $K^\prime$ form a group, but only under the condition that
 $X^\prime$ is complex. To match the reality condition for $X^\prime$ with
 that of $X$ in eq.(4) and to conserve the group properties of (6) one
 should represent $X^\prime$ in the form
 $$
i X^\prime=iX+{1\over 2}\theta\bar\theta.
$$
The resulting expression for the super--Poincar\'e  group is as follows
\begin{equation}\label{6}
{G_{SP}=
\left(\begin{array}{ccc}
1&\theta&iX+{1\over 2}\theta\bar\theta\\
0&1&\bar\theta\\
0&0&1
\end{array}\right)\times
\left(\begin{array}{ccc}
L&0&0 \\
0&1&0 \\
0&0&(L^+)^{-1}
\end{array}\right)}.
\end{equation}
In \cite{2,3} more general case of N--extended super--Poincar\'e
group was considered. In this case $\theta$--spinors acquire additional
internal symmetry index and, in the second factor in (7) the unit matrix
is replaced by an internal symmetry group matrix.

 From (7) one gets the transformation law for the superspace coordinates:
\begin{equation}\label{7}
{ X^\prime=X+i\bar\theta\gamma \epsilon,}
{}~~~{\theta^\prime=\theta+\epsilon,~~~{\bar\theta}^\prime
=\bar\theta+\bar\epsilon}
\end{equation}
as well as the following expressions for the left--invariant
superspace vielbein one--forms
\begin{equation}\label{8a}
{ e^a=dX^a-i\bar\theta\gamma^ad\theta}
\end{equation}
\begin{equation}\label{8b}
\bf{e^\alpha=d\theta^\alpha}.
\end{equation}
The latter are obtained as components of $K^{-1}dK$ corresponding to the
generators of $K$ now being the supertranslation subgroup of (7).

The action for the Goldstone fermions is the pullback of a target
superspace
differential
four--form onto the four--dimensional Minkovski subspace (the world space)
\begin{equation}\label{9}
{S={1\over {24}}\varepsilon_{absd}e^ae^be^ce^d}.
\end{equation}
In eq. (\ref{9}) as well as in further formulas exterior product
is implied.

The action (\ref{9}) gives the first example of a method widely applied
nowadays for constructing superparticle, superstring and supermembrane
actions out of the target superspace vielbein forms with their pulling
back onto the worldsheet.

 Now we can see what are the differences between
(\ref{9}) and (\ref{01}).  Firstly, the Lagrangian (\ref{9}) is
constructed not out of the 1--forms (\ref{8b}), as it may be expected,
but out of the 1--forms (\ref{8a}).
Secondly, the Lagrangian (\ref{9}) is not the metric of a Riemann space as
in (\ref{01}), but the 4--volume form.

Interactions of the Goldstone fermions with particles of spin 0,
${1\over2}$ and 1 were considered in \cite{3,26}
with the use of the method developed in
  \cite{24,25}. Notwithstanding the differences from the internal symmetry
  case mentioned above, all soft Goldstone particle theorems were
  reproduced, and we became sure that our generalization of the Poincar\'e
  group may have something to do with reality. Our first paper was
  entitled ``Is the neutrino a Goldstone particle?'', but since gauging
  the Poincar\'e group is directly connected with the Einstein--Cartan
  theory of gravity, the same should take place for any generalization of
  the Poincar\'e group.  Therefore, our paper was concluded by the
   following sentence
   \begin{quotation} `{\sl...the gravitational interaction
  may be included by means of introducing the gauge fields for the
  Poincar\'e group. Note that if the gauge field for the fermionic
  transformation is also introduced, then as a result of the Higgs effect
  the massive gauge field with spin three--halves appears and the
  considered Goldstone particle with spin one--half disappears...}'
  \end{quotation}
  which was the first may be somewhat implicit mentioning of the theory of
  supergravity as a theory containing the Einstein--Cartan action
  and the action for the Rarita--Schwinger field, the latter being
  massless in the absence of the spontaneous supersymmetry breaking.
    We also considered the possibility that the
  super--Poincar\'e group might be an approximate symmetry so that (in
  analogy with $\pi$, $\rho$ and $A^\prime$ mesons in hadron physics
\cite{14}) Goldstone particles with nonzero mass and the Higgs effect
coexisted.

\section{Gauged super--Poincar\'e group}
Now let us go over to the procedure of gauging super--Poincar\'e group.

Supersymmetry is spontaneously broken symmetry. Therefore, it is
convenient to consider its gauged version from the very beginning in the
form containing Goldstone fermion fields.

Note that the 1--forms (\ref{8a}) as well as (\ref{8b}) are not covariant
with respect to the local transformations of the group under
consideration. But if local transformation laws of Goldstone and gauge
fields {\sl are correlated}
$${K^\prime=G_LKH},$$
$${A^\prime(d)=G_LA(d)G^{-1}_L+G_LdG^{-1}_L},$$
the sum
$${\tilde A(d)=K^{-1}dK+K^{-1}A(d)K}$$
transforms as
$${\tilde A^\prime(d)=H\tilde{A}(d)H^{-1}+HdH^{-1}}.$$
Only $\bf{<\tilde A(d)>_h}$ are the gauge fields, and
$\bf{<\tilde A(d)>_k}$ are non--gauge fields which have absorbed the
Goldstone degrees of freedom and transform covariantly under subgroup H.

 The differential forms presented above can be used as blocks for
constructing invariants of the local group,  and then summing up all
independent invariants one gets the general form of the gauge field
action. It is self--evident that the action constructed this way contains
not only the terms with spontaneously broken symmetry but a gauge
invariant action as well. The latter can be
easily extracted by counting the physical degrees of freedom. In other
words, one should recall that the basic essence of the Higgs effect is
that the Goldstone degrees of freedom though absorbed by the gauge fields
are physical degrees of freedom. If some combination of the local
invariants considered as an action results in two
physical gauge field degrees of
freedom, it means that corresponding Goldstone fields do not contribute
and they either drop out of the action or become auxiliary fields. The
latter, as shown in the next section, is the case for N=1 supergravity.

Here we remind what methods for considering the Higgs effect in the
case of internal symmetry groups were used at the end of 60's and at
the beginning of 70's. Such reminding is useful for the reader to
better understand the state of the development of theoretical physics at
that time, and, as it has been already said, the consideration of internal
symmetries and supersymmetry in parallel helps one to formulate and to
solve the problems in the latter case.

In the case of internal symmetry groups the approach for considering the
Higgs effect may be roughly divided on to the following stages
\begin{itemize}
\item
the method described above for constructing the action out of the
differential forms with the correlated transformation law of the gauge and
Goldstone fields. The advantage of this method is its generality, the
drawback is its phenomenological nature which exhibits itself in the
presence of a number of arbitrary constants;
\item
studying simplified models constructed for understanding possible
mechanisms of spontaneous symmetry breakdown;
\item
studying realistic gauge field theories coupled to matter fields which can
ensure spontaneous symmetry breaking.
\end{itemize}

All these stages were applied and are being applied now to supersymmetric
field theories.

The gauge fields for the N=1 super--Poincar\'e group are described by a
$g_{susy}$--valued matrix
$$
A_{SUSY}(d)=\left(
\begin{array}{ccc}
\omega(d)&\psi(d)&e(d)\\
0        &  0    &\bar\psi(d)\\
0        &  0    &\bar\omega(d)\\
\end{array}\right)
$$
where
${\omega(d)}$ is the Lorentz connection,
${\psi(d)}$ is the Rarita--Schwin\-ger gauge field and
${e(d)}$ is the vierbein 1--form.

Corresponding ${\tilde A_{SUSY}(d)}$ forms are
$${
\tilde e(d)=e(d)+DX
+i\left[(2\psi(d)+
D\theta)\bar\theta-\theta(2\bar\psi(d)+D\bar\theta)\right],}
$$
\begin{equation}\label{forms}
\tilde\psi(d)=\psi(d)+D\theta,~~~
\tilde\omega(d)=\omega(d),
\end{equation}
$${
\tilde R(d,d^\prime)=R(d,d^\prime)=d\omega(d^\prime)-d^\prime\omega(d)+
\left[\omega(d),\omega(d^\prime)\right].}
$$
where $D$ is the covariant differential with the connection form
$\omega (d)$

\section{Supergravity action and the super--Higgs effect}
Contracting the indices of the differential forms (\ref{forms})
one can construct the following locally invariant differential
four--forms \cite{7,8}:

\bigskip
${ W_1=\tilde R(d_1,d_2)\tilde e(d_3)\tilde e(d_4)}$~--~ Einstein--Cartan
action

\bigskip
${
W_2=D\bar{\tilde\psi}(d_1,d_2)\tilde e(d_3)\tilde \psi(d_4)}$
{}~--~Rarita--Schwinger kinetic~term

\bigskip
${
W_3=\tilde e(d_1)\tilde e(d_2)\tilde e(d_3)\tilde e(d_4)}$~~--~~
cosmological term

\bigskip
${
W_4=\bar{\tilde\psi}(d_1)\tilde e(d_2)\tilde e(d_3)\tilde \psi(d_4)}$
{}~--~Rarita--Schwinger mass~term

The resulting action for N=1 supergravity is the sum
\begin{equation}\label{act}
S=S_1+S_2
\end{equation}

with
\begin{equation}\label{s1}
S_1=a_1W_1+a_2W_2
\end{equation}
being the pure SUGRA action and
\begin{equation}\label{s2}
S_2=a_3W_3+a_4W_4
\end{equation}
being the terms which arise due to the spontaneous
breakdown of the super--Poincar\'e group. Let us stress once again that
each of the terms $W_i$ and, hence, action $S_1$ and $S_2$ obey local
supersymmetry.

The action $S_1$ (\ref{s1}) is the same as one which is now accepted as the
action of  $N=1$ supergravity, and
since the action for any newly proposed dynamical system completely
determines all its physical properties, the time, when the action was
firstly written down is the date of the discovery of the theory. So N=1
supergravity as a physical theory was discovered in 1973. At the XVII
International Conference on High Energy Physics (London, 1974) B. Zumino
told \cite{39}:
\begin{quotation}
`{\sl Volkov and Soroka \cite{2} have developed a description of curved
superspace which combined gravitational theory with interaction of
particles of spin 3/2, 1 and 1/2. Can a theory of this kind, because of
compensation of divergences due to supersymmetry, provide a
renormalizable description of gravitational interaction?}'
\end{quotation}

Now we proceed to explain (13-15) by counting the physical degrees of
freedom of the gauge fields. It is self--evident that due to the fact that
all terms are differential 4--forms the gravitational field has two
physical
components and corresponds to the unbroken symmetry. The Rarita--Schwinger
field corresponds to the broken or unbroken local supersymmetry if it has
four or two physical
degrees of freedom, respectively. Thus, using this simple
argument of counting the degrees of freedom one can find out whether
supersymmetry is broken or not. This depends on the values of the
coefficients $a_i$ in (\ref{act}). To determine the coefficients it is
sufficient to consider particular cases, the simplest examples
being the Minkovski and the anti--de--Sitter space (if the mass term with
a definite ratio of the mass to the curvature of the anti--de--Sitter
space is added) as backgrounds for the Rarita-Schwinger field, and then
go over to more general cases. It can be
shown quite easily that if a background Einstein field satisfies the
equation of motion for the term $W_1$ the Rarita--Schwinger field, in this
background, has only two physical components, the Goldstone fermions do
not contribute to the first two terms in the sum (\ref{act}) and all
effects of the spontaneous breaking of the super--Poincar\'e group are
contained only in the third and the fourth term. If in (\ref{act})
$a_i$ are arbitrary the Rarita--Schwinger field has four degrees of
freedom, so in the general case the Goldstone fermions are physical
degrees of freedom, which is the realization of the super--Higgs effect.


The auxiliary Goldstone fermion fields (as well as $X^a$) can be excluded
from the action $S_1$ without changing its physical contents.
But, firstly, this procedure is not unique, and, secondly, which is more
important, the off--mass--shell local invariance of $S_1$ is lost and
reduced to the local invariance on the mass shell with all its unpleasant
consequences that the algebra is not closed, structure ``constants''
depend on fields etc. And since on the mass shell the Lorentz
connection $\omega(d)$ becomes a function of $e(d)$ and $\psi(d)$ its
transformation law changes and becomes rather complicated. The action
$S_1$, with the local supersymmetry held on the mass shell, was regained
in 1976 \cite{28,29}. In the review \cite{ferr} (p.319) the main result of
\cite{29} was formulated as follows:
\begin{quotation}
`{\sl Deser and Zumino \cite{29} have
shown that supergravity is nothing but the spin $3/2$ Rarita -- Schwinger
Lagrangian minimally coupled to Cartan first order formalism of general
relativity}.'
\end{quotation}

As to the terms $S_2$ (\ref{s2}) in (\ref{act}), they are reproduced in
all N=1 supergravity theories with matter fields specifically added to
make supersymmetry spontaneously broken. Of course, in such theories  the
constants $a_3$ and $a_4$ are determined by the choice of the matter
fields, their masses and interaction constants.

There is a number of papers where the equivalence of our approach to
the conventional one is established. See, for example, \cite{100} and
references therein.

Our approach can be applied to any $N \leq 8$ case as well.
In papers \cite{7,8} we wrote that we considered ``the simplest
possible local invariants''. In $N=1$ case the local invariants $W_i$
constitute the complete set. When $N$  increases the
number of local invariants also increases. If $N > 2$  not only gauge and
Goldstone fields, but also all non--gauge fields which are the
superpartners of the gauge fields in the supergravity multiplet, should be
used for the construction of the complete set of invariants. The situation
is the same as in the case of spontaneously broken internal symmetry with
the only difference that in the case of supersymmetry gauge and non--gauge
fields may belong to the same supermultiplet.

\section{Superspace formulation of Supergravity (First steps into
Superspace)}

The first superspace formulation of supergravity was attempted by
R. Arnowitt, P. Nath and B. Zumino \cite{30}. They proposed the
generalization of the Einstein 4--dimensional action to superspace with
coordinates $(x,\theta)$ in the following form
$$ S=\int R\sqrt{Ber~g}
d^4\theta d^4x.
$$
The analysis of this action revealed that it contains
a number of drawbacks such as
\begin{description}
\item[a)] the presence
of fermionic ghosts,
\item[b)] the action does not admit ``flat''
superspace as a solution;
\item[c)] for any solution torsion is zero,
which is a consequence of Riemann geometry chosen;
\item[d)] the holonomy
group of the tangent space was OSp(3,1/4) which is too large.
\end{description}

Upon realizing these drawbacks  two groups of authors \cite{31,32}
generalized the E. Cartan's method of differential geometry to superspace
and argued that \begin{itemize} \item the holonomy group for superspace
curvature is the Lorentz group; \item  flat superspace has torsion and,
hence, torsion has to be included into the theory of supergravity.
\end{itemize}

These points are now part of all existing versions of supergravity.
The intensive development of the superspace approach to supergravity has
been carried out starting from 1976 and up to now. At the first stages
Wess and Zumino, Ogievetski and Sokatchev, Siegel, Ferrara and Ro\'cek,
Scherk and many others have made significant contribution to this
development.

\section{Resum\'e}

Before 1976 the following general ideas had been proposed and the
following results obtained
\begin{itemize}
\item
Super--Poincar\'e group (N=1, and ${\bf N>1}$)
\item
Flat superspace, supervielbeins.
\item
Notion of Goldstone fermions
\item
Phenomenological Lagrangian for Goldstone
fer\-mi\-ons interacting
with particles of s=0, ${\bf 1\over 2}$  and 1
\item
Gauging the Super--Poincar\'e group
\item
Supergravity action, N=1 (the off--mass--shell formulation).
\item
Super--Higgs effect, ${\bf N\ge 1}$
\item
First attempts to the superspace formulation of supergravity.
\end{itemize}

As concerns the techniques developed, it is mostly connected with the
application of E. Cartan's methods of the exterior differential forms and
differential geometry. E. Cartan's methods were applied to the component
formalism for constructing the Goldstone fermion and supergravity action,
and to trace a way to the superspace formulation of supergravity.

Nowadays E. Cartan's methods, being the fundamentals of the fiber bundle
theory, are among the main tools used in theoretical physics.

\bigskip
\bigskip
\begin{center}
\underline{~~~~~~~~~~~~~~~~~~~~~~~~~~~~~~~~~~~~~~~~~~~~~~~~~~~}
\end{center}

As the speakers at the Conference were asked to present not only their
ideas and results but also to describe the path along which they went to
reach them, I shall describe some of the main impulses and motivation which
helped my coauthors and me to pass the way to supersymmetry and
supergravity.

Of the greatest importance was, as I call it, my personal inclination to
the problem of connection between the spin and statistics, and to related
topics. My first papers had been done in that direction of
research.

My PhD thesis was devoted to the calculation of radiative corrections to
some effects in the quantum electrodynamics of scalar particles. I had
performed it following J. Schwinger papers, and from that time I started to
consider him as my teacher by correspondence and, as his student by
correspondence, I carefully studied all his papers  at stock.

On my only personal meeting with J. Schwinger in Kiev in 1959 we had a
fruitful discussion during which I proposed a possible generalization of
an admissible class of variations in  Schwinger's quantum dynamical
principle by connecting the variations with the symmetry properties of the
action. A little bit later, J. Schwinger published a paper \cite{33} in
which the generalization proposed was elaborated. For me this episode
became as an exam which I successfully passed to my teacher by
correspondence.  Listening the talk by S. Glashow at the Erice Conference
I envied  his lucky fortune to be in everyday contact with J. Schwinger.

 Of course, the use of Grassmann variables as
variations of fermion fields in the Schwin\-ger quantum variation
principle was an essential step to supersymmetry, while the
Rarita--Schwin\-ger field was a step to supergravity. I am sure that J.
Schwinger had been quite prepared to write down the supergravity action in
the form of (\ref{act}) in the middle of 60's or even earlier, and
only his engaging in studying other problems can explain why he had not
done it.

At the end of 50's W. Heizenberg and W. Pauli proposed a
 non--linear theory which claimed to become the base for understanding the
 spectra and interactions of elementary particles. A little bit later, W.
 Pauli changed his role as the coauthor for the role of the most severe
 Heizenberg's opponents\footnote{W. Heizenberg describes this period in
 a monograph \cite{34}}.  But W.  Heizenberg continued his work, and,
 trying to overcome the mathematical difficulties of the theory, used his
 great intuition to find the place for each of the existing particles in
 the scheme. He succeeded in this attempt for a number of particles.
 Considering the neutrino, W.  Heizenberg proposed that it might be a
 Goldstone particle connected with the spontaneous breakdown of parity
 \cite{35}.  The assumption impressed me immensely, but at that time I had
 not been prepared to implement this idea.

 As I mentioned in section 2 my approach to the
 Phenomenological Lagrangians was different in some points from other
 approaches. The difference was that the starting point of \cite{22} was
 the action (\ref{01}) with an arbitrary metric tensor on which no
 symmetry requirements were imposed. Calculating the on--mass--shell
 scattering amplitudes I got that they depend only on manifestly covariant
 quantities such as the curvature tensor and its covariant derivatives.
 Upon imposing the simplest of possible restrictions, namely, that the
 covariant derivative of the curvature tensor vanishes $$ R_{absd,f}=0 $$
 I became
 aware that this condition is the definition of the symmetric spaces
 proposed by E.  Cartan. I began to study his works and they  opened for
 me the mathematical beauty of differential geometry and taught many
 lessons on how one might apply it to physical theories.

 In previous sections I have stressed the role of Goldstone, Nambu,
 Schwinger, Weinberg, Higgs and others whose papers introduced me into the
 world of nonlinear realized (secret) symmetries.

 So the way to the discovery of supersymmetry and supergravity was rather
 long and not a straight line. And now I see that the seeds of
 supersymmetry had been thrown onto the fertile soil and beautiful
 mathematical structures had grown out of them, and all of us, theorists
 and experimentalists, are waiting for the time when we can touch and taste
 the fruit.

 \bigskip
 \bigskip

 I am very grateful to the Organizing Committee of the Conference
 for giving me this unique opportunity to illuminate the period
 in the development of supergravity which is usually omitted
 exposing the history of its discovery.

 \end{document}